# Experimental observation of optical vortex evolution in a Gaussian beam with an embedded fractional phase step

W.M. Lee[1], X.-C. Yuan[1]* and K. Dholakia[2]

[1]Photonics Research Center, School of Electrical and Electronic Engineering, Nanyang Technological University, Nanyang Avenue, Singapore 639798

[2]School of Physics and Astronomy, University of St. Andrews, North Haugh, St. Andrews, Fife KY16 9SS, Scotland

Laguerre-Gaussian beams may possess an azimuthal phase variation of $2\pi l$, where $l$ is an integer, around the beam axis, resulting in an annular intensity distribution. This azimuthal phase variation and associated vortices can be visualized through the appearance of forked fringes, when interfering the LG beam with its mirror image (a beam of opposite helicity) at an angle. In this paper, we examine the evolution of optical beams with a fractional phase step hosted within a Gaussian beam by experimental analysis of both the phase and intensity distribution. To generate these beams, we introduce differing fractional (non-integer) topological charge variations within a Gaussian beam generated using a spatial light modulator (SLM). We detect the evolution of the vortex from the increase of the fractional phase step by interfering two beams of opposite but equal fractional phase step increment. The interference pattern generated shows evidence of the birth of an additional single extra charge as the fractional phase step increase extends above a half –integer value.

Dislocations or phase defects within a monochromatic wave have been analyzed in detail by J F .Nye *et al* [1]. In the analysis, a number of optical wavefront dislocations have been identified including screw dislocations, edge dislocations and mixed screw-edge dislocations. The most commonly studied phase defect in an optical field is screw-dislocation. A screw dislocation in an optical beam results in a phase singularity at the centre of the beam where the phase is not defined. Hence when a screw dislocation is hosted within a Gaussian beam, the resulting beam will exhibit an annular intensity profile whilst maintaining a helical phase structure. A typical example of such a light field is a Laguerre-Gaussian (LG) beam [2].

The phase singularity and orbital angular momentum embedded in the LG beam is due to its helical phase structure, denoted in the mode description by $e^{il\phi}$ where $2\pi l$ is the helical phase around the circumference of the optical field [2]. Such beams have generated widespread interest in the last decade for optical rotation of micro-particles [3], quantum entanglement of photons [4], nonlinear optics [5] and optical vortex interactions [6].

Recently there has been significant interest in fractional half-charge dislocations embedded within optical beams [7 - 12]. To date, for the generation of an optical beam of half-integer fractional charge, both spiral phase plate [10, 11] and off-axis holograms [7, 8] have been used. I.V Basistiy *et al* made use of a computer-generated hologram, based on a half-integer screw dislocation, to generate a monochromatic beam. In a recent publication, Berry [12] discusses mathematically the evolution of waves with phase singularities of $2\pi l$, where $l$ may now be either integer or non-integer. For the case of the fractional phase step ($2\pi l$) being non-integer, a key theoretical prediction of Berry is the birth of a vortex within the beam as the fractional phase step reaches and passes a half-integer value. Thus to experimentally investigate such fractional topological charges we need to both generalize the range of fractional phase steps available as well as find a way of studying the evolution of the beam as this fractional phase step is varied .

In this paper, we investigate fractional phase steps hosted in a Gaussian beam, by means of a spatial light modulator (SLM). In contrast to previous studies we are able to generate any arbitrary fractional phase step within the light field and specifically visualize the birth and evolution of a vortex as we increase the fractional order of the light field. Detection of the birth of this vortex is achieved through an interference [13] experiment that very clearly elucidates the evolution of the topological charge within the beam profile. Such dynamic generation of the fractional phase step provides a detailed study of the fractional phase shift within the optical beam. In comparison to static off-axis holograms and spiral phase plates, the SLM generation and subsequent interferometric analysis is able to provide a real-time study of such phase evolution. In this instance this permits us to experimentally verify one of the key predictions of Berry's work with regard to beams with fractional (non-integer) phase steps.

LG beams may be generated by off-axis computer generated holograms (CGH) with dislocations. This is based on the fact that when a plane wave interferes at an angle with a Laguerre-Gaussian beam, shown in Equation (1), of the following expression, a dislocation is formed. Such holograms with dislocations appear in the form of a forked diffraction grating [14]:

$$E(LG_p^l) \propto \left[\frac{r\sqrt{2}}{w(z)}\right]^{|l|} L_p^{|l|}\left[\frac{2r^2}{w^2(z)}\right] \cdot e^{\left(\frac{-r^2}{w^2(z)}\right)} \cdot e^{\left(\frac{-ikr^2 z}{2(z^2+z_R^2)}\right)} \cdot e^{(-il\phi)} \cdot e^{-i(2p+|l|+1)\tan^{-1}\frac{z}{z_R}} \quad (1)$$

where $L_p^{|l|}$ is a generalized associated Laguerre polynomial, w(z) is the radius of the beam at the position z, $z_R$ is the Rayleigh range.

The reconstructed optical beam $E(r,\phi,z)$, possess a phase function of $e^{(-il\phi)}$, from the computer generated hologram in the diffractive orders can be simply approximated to as shown in Equation (2),

$$E(r,\phi,z) \propto e^{(\frac{-r^2}{w^2(z)})} \cdot e^{(-il\phi)} \qquad (2)$$

where *w(z)* is the radius of the beam at the position z and $\phi$ is arctan(y/x). Using this relation, the fractional phase azimuthal variation hosted within a Gaussian beam may be given by Equation (3).

$$E(r,\phi,z) \propto e^{(\frac{-r^2}{w^2(z)})} \cdot e^{(-il_1\phi)} \cdot e^{(-i\frac{l_2}{10}\phi)} \qquad (3)$$

In this experimental work, we generated beams as in equation 3 choosing to implement a fractional phase azimuthal variation increasing in steps of 0.1 where $l_1$ denotes the starting integer value of the phase step and $l_2/10$ ($l_2$ integer) is the fractional increment we add to the topological charge. Thus in these cases, $l_1$=0, 1, 2, 3, 4… but $l_2$ varies only from 1 to 10 for each integer value of $l_1$.

Our experimental setup consists of a collimated laser beam of 632.8nm (maximum power 30mW) which is directed onto the SLM. The SLM is from the Boulder Nonlinear Systems (BNS), which is a nematic type working on reflection, containing have more than 64 phase levels and each pixel size of the SLM is 15 μm by 15 μm. The loaded grayscale sinusoidal off-axis hologram, contain vertical gratings as shown in Figure.1, have 512 by 512 sampling points. The output consists of optical beams with a fractional phase steps in the 1st diffraction orders directed in the horizontal manner. The 1$^{st}$ order beam is then recorded with "Beamstar-1500" CCD beam profiler. By interfering the generated 1$^{st}$ order diffracting beam hosting the fractional phase azimuthal variation with a plane wave mathematically, a series of off-axis computer generated holograms are generated on an SLM as shown in Figure 1. In Figure 1, it is interesting to note that as $l_2$ varies only from 1 to 10 for $l_1$=0, the centre of the CGH generated slowly shifted to the left by an interval of $\pi/5$. When $l_2$ reaches 10, a charge 1 dislocation is formed at the centre of the CGH, which results in a single integer phase step embedded within the charge one LG beam. Hence the fractional phase increment from also be observed through the changes in the CGH.

By observing the generated fractional phase step within the beam as shown in Figure 2 and 3, we can clearly observe the changes in the optical wavefront at a distance of 0.7$Z_R$, where $Z_R$ is Rayleigh range of the beam. In our experiment, we were able to observe the appearance of a step discontinuity , for $l_2 \neq 0$, in the wavefront modulation through its intensity pattern as shown in Figure 2 and 3. Our experimental results for the intensity distribution shows that even a fractional phase step of 0.1 hosted with optical beam will cause a defect to exist within the optical beam. This defect is solely due to the presence of the fractional phase step within the optical beam.

It is also possible for us to consider an optical beam of fractional phase step as a superposition of several axially symmetric optical beams of different integer topological charges (Laguerre-Gaussian beams) [7, 9]. It is well known from the work of Beijersbergen *et al* [9] that to obtain the accurate mode description of the output of a spiral phase plate or hologram we may decompose the beam in terms of the full orthonormal basis set of LG beams giving us the appropriate weightings, for each constituent LG mode. Modes of fractional order require significant weightings, in principle of an infinite number of modes, and it has been suggested they can be used for more detailed studies of high-dimensional quantum entanglement and violations of Bell's inequalities [11].

By interfering the opposing diffractive orders (±1) from the SLM, we are able to show the evolution of the beam through the fork fringe formation: the birth of a new vortex manifests itself as the clear appearance of a forked fringe in the aforementioned interference pattern

In the experimental set-up (described previously in [13]) we interfere both the +1$^{st}$ and -1$^{st}$ diffraction order from the off-axis dynamic hologram using the two mirrors. By interfering the two opposite diffraction orders at an angle, we can observe the helicity of the individual optical beam and obviate any errors due to Gouy phase mismatch [15]. Hence, the resulting helicity or fork fringes $l_R$ will follow an express of the form:

$$l_R = \left| l_{LG_0^l} - l_{LG_0^{-l}} \right| \tag{5}$$

As shown theoretically by Berry [12], when the fractional phase step increases beyond the half-integer value, an extra vortex is born within the optical beam.

$$l_{LG_0^{l_2}} = \text{int}\left(\frac{l_2}{10} + \frac{1}{2}\right) \tag{6}$$

where $l_2$ is a varies from 1 to 10 in this case.

We conclusively verify this in our experiment, as we are able to observe such evolution of the extra vortex as seen explicitly in figures 5 and 6. In Figure 5, the fringes that we observe for the fractional azimuthal phase increment are clearly observed as $l_2$ /10 increases from 0.1 to 1. The key result is that the forked fringes appear (a new vortex is born) when the fractional topological charge is 0.5 and we see the formation of a charge 2 fork fringe, which equals to |1-(-

1)| using Equation 5. In Figure 6, there is also a clear evolution of an extra fringe at the fractional phase step of 1.5., again demonstrating the birth of new vortex within the beam.

On top of the observation at the distance below that of Rayleigh's length of $Z_R$, we also observed that the optical beam embedded with a half-integer phase step will possess an integer-based vortex when propagating to the far-field, as shown in Figure.7, which has also been experimentally observed by Basistiy [8, 9]. Hence this reaffirms the point that once the vortex is born in the near-field, the optical vortex is structural stable upon propagation towards the far-field. However, the different is that the position of the vortex moves from off the beam axis onto the beam central axis as the phase step is incremented to the next integer phase step.

In conclusion, we have experimentally generated optical beams with arbritary azimuthal fractional phase step increments using an SLM. We have shown the evolution of the fractional phase variation through a Michelson interferometer. These changes in the fractional phase variation show the birth of new vortex as the fractional phase steps reaches and passes the half-integer value. The interferometric patterns show how an additional vortex is created (born) and evolves within the beam. Fractional phase steps are of interest in advanced studies of quantum entanglement [11] and have analogies to quantum flux in the Aharonov-Bohm effect [12].

Kishan Dholakia acknowledges the support of a Tan Chin Tuan Exchange Fellowship.

Figure Caption

Figure 1 –CGHs for optical beam with varying fractional phase azimuthal, where $l = l_1 + l_2/10$.

Figure 2 – Experimental result of LG - $l$ =0 to 1.0, where $l = l_1 + l_2/10$ taken at a propagation distance of 0.7$z_R$

Figure 3 – Experimental result of Opposite charge LG - $l$ =1 to 2, where $l = l_1 + l_2/10$ taken at a propagation distance of 0.7$z_R$.

Figure 4 – Optical set-up to obtain the fractional optical beam using the SLM and the interference of the beams to reveal the azimuthal phase variation. BS denotess the non-polarizing beam-splitter, M is the reflective mirror and CCD is a "Beamstar-1500" CCD beam profiler used for beam analysis..

Figure 5 – Experimental results of LG beam interfering with opposite helicity at a distance of 0.5$z_R$ - $l$ =0.1 to 1.0, where $l = l_1 + l_2/10$ (The red circle indicate the formation of the extra fork fringe wit (additional vortex formation)

Figure 6 – Experimental results of LG beam interfering with opposite helicity at a distance of 0.7$z_R$ - $l$ =1 to 2.0, where $l = l_1 + l_2/10$ (The red circle indicate the formation of the extra fork fringe with respect to the extra single vortex formation)

Figure 7 – Experimental results of zeroth order diffractive order with the 1$^{st}$ order diffractive order of varying phase step at a distance of $z_R$. (The red circle indicate the formation of the integer vortex)

> A- the interference of the zeroth order diffractive order with the 1$^{st}$ order diffractive order with zero topological charge where there is no dislocation detected.
>
> B - the interference of the zeroth order diffractive order with the 1$^{st}$ order diffractive order with fractional topological charge of $l$ =0.5 where there is a single off-axis dislocation detected.
>
> C - the interference of the zeroth order diffractive order with the 1$^{st}$ order diffractive order with fractional topological charge of $l$ = 1 where there is a single on-axis dislocation detected.

Figure 1

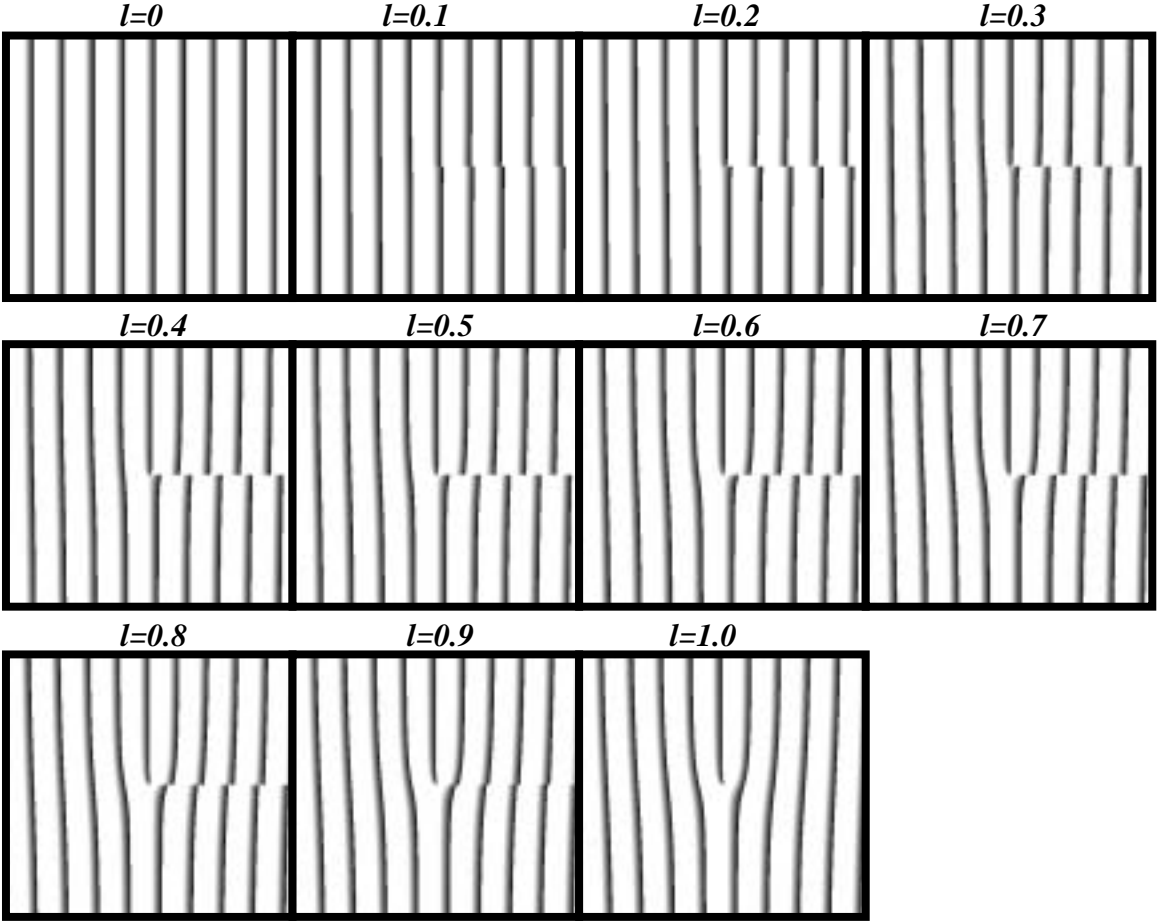

Figure 2

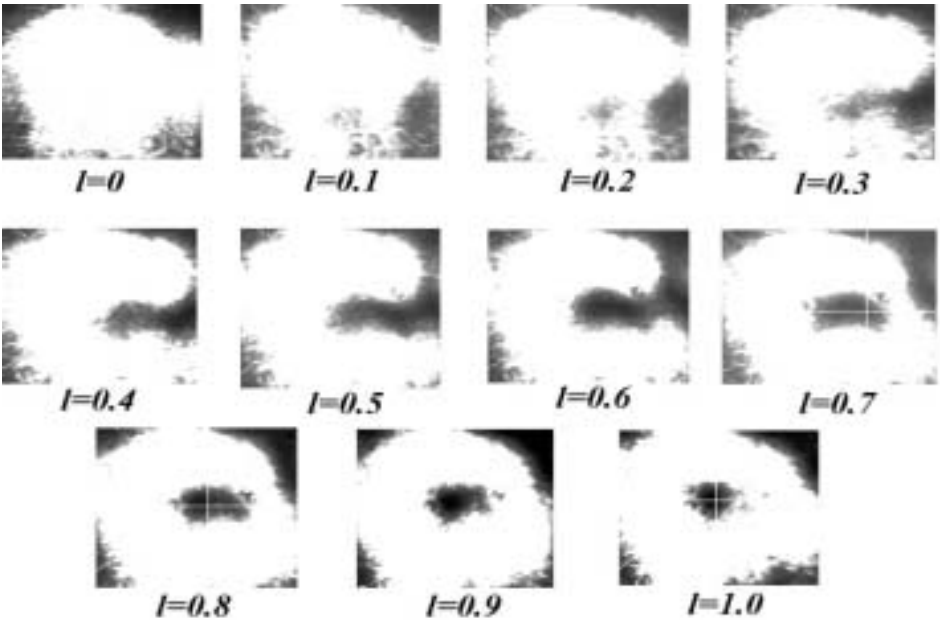

Figure.3

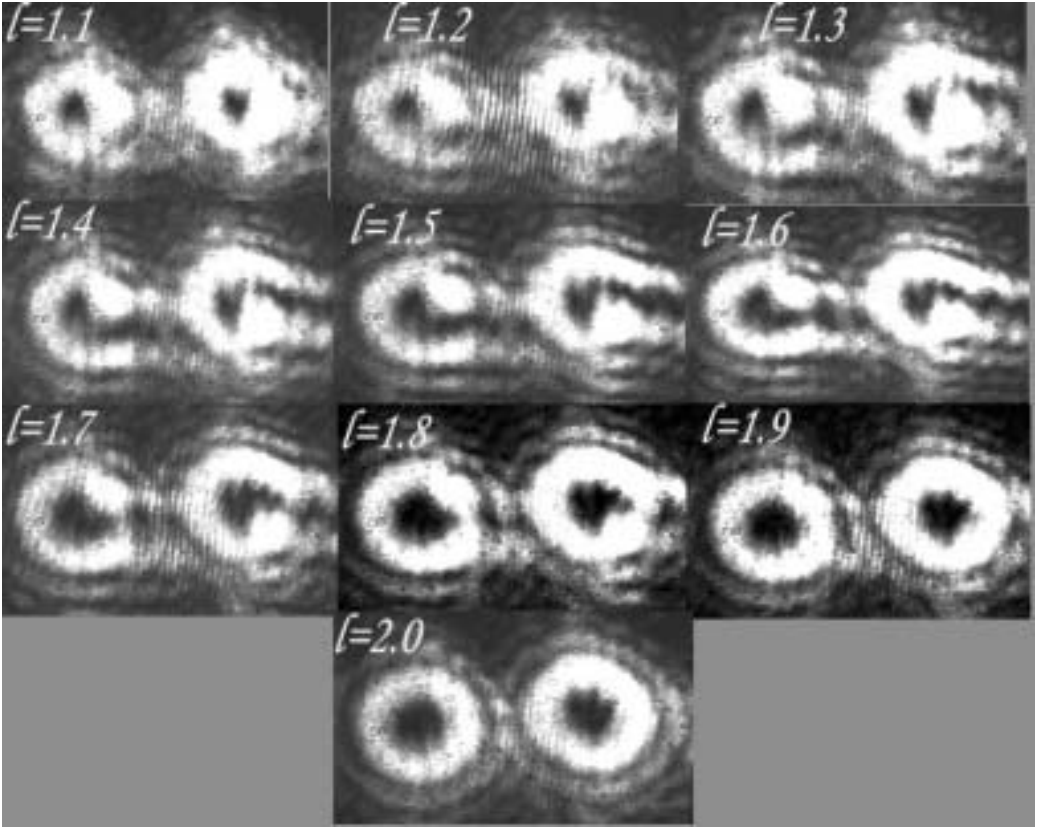

Figure.4

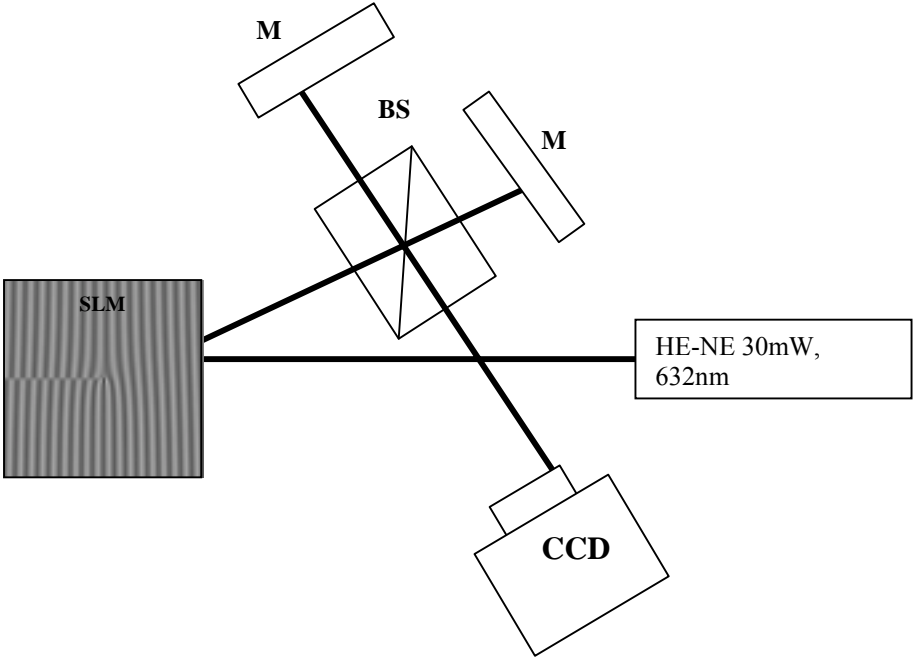

Figure.5

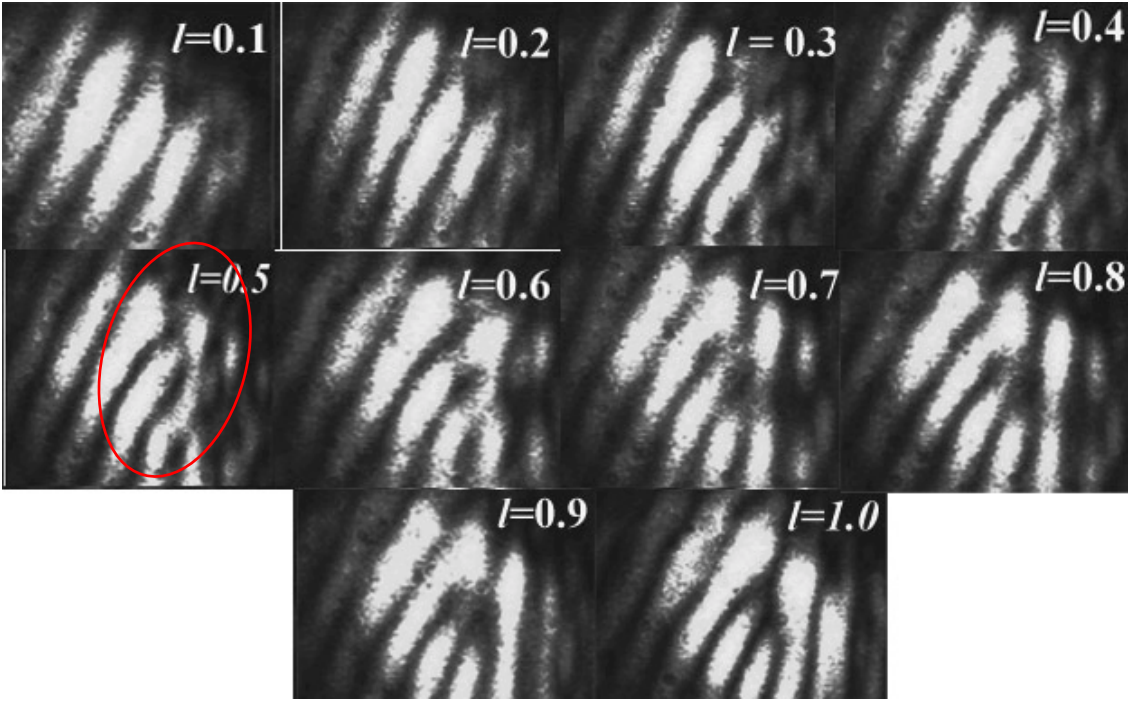

Figure.6

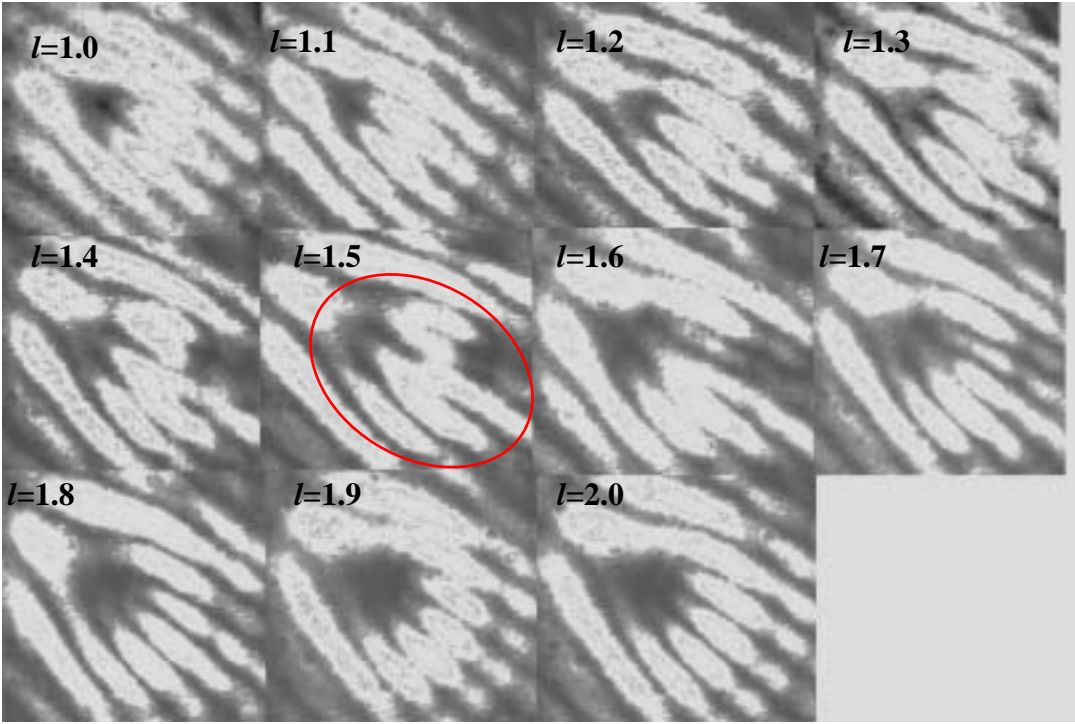

Figure.7

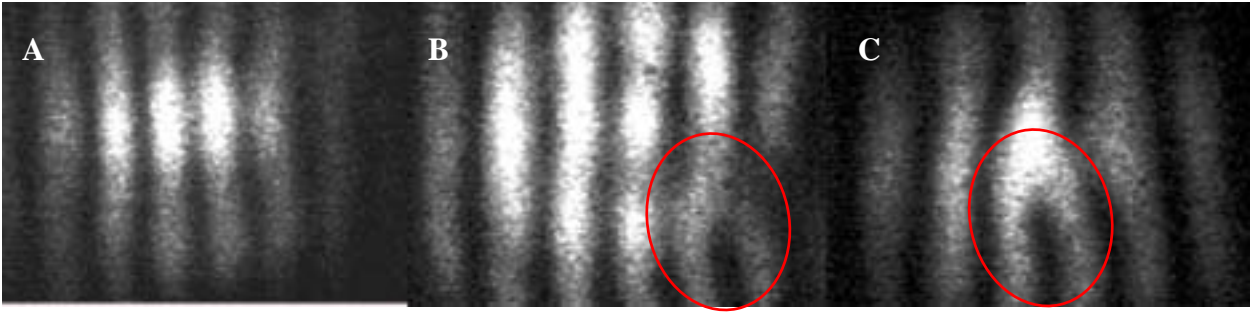